\documentclass[12pt]{elsarticle}

\usepackage{epsfig}
\usepackage{amsmath}
\usepackage{amsfonts}
\usepackage{amssymb}
\newcommand{\beq}{\begin{equation}}
\newcommand{\eeq}{\end{equation}}
\newcommand{\bea}{\begin{eqnarray}}
\newcommand{\eea}{\end{eqnarray}}
\newcommand{\hf} {\frac{1}{2}}
\newcommand{\nn}{\nonumber\\}

\newcommand\eqn[1]{Eq.\,(\ref{#1})}

\newcommand\fig[1]{Fig.\,{\ref{#1}}}
\newcommand\sect[1]{Sect.\,{\ref{#1}}}

\def\mr#1{{\mathrm{#1}}}

\def\tr{\mbox{tr}}
\def\Tr{\mbox{Tr}}
\def\dblone{\hbox{$1\hskip -1.2pt\vrule depth 0pt height 1.6ex width 0.7pt
\vrule depth 0pt height 0.3pt width 0.12em$}}

\begin{document}

\title{Degeneracy induced scaling of the correlation length for periodic models}

\author{S. Nagy}
\address{Department of Theoretical Physics, University of Debrecen,
P.O. Box 5, H-4010 Debrecen, Hungary}
\address{MTA-DE Research Group in Particle Physics, H-4010 Debrecen P.O.Box 105, Hungary}

\begin{abstract}
The broken symmetric phase of scalar models exhibits an infrared fixed point
which is induced by the degenerate effective potential. The definition of the
correlation length in the infrared regime enables us to determine
the type of the phase transition in the model. It is shown that the massive
sine-Gordon model exhibits a continuous, while the layered sine-Gordon model
has an infinite order Kosterlitz-Thouless type phase transition.
\end{abstract}

\maketitle

\section{Introduction}\label{sec:intro}

The renormalization group (RG) method provides a powerful tool to map the phase
structure of field theoretical models \cite{Wetterich,effrg}.
The phase space usually contains singularity points,
which correspond to stationary solutions of the RG flow equations.
The linearization of the RG equation in their vicinity results in the power-law scaling of
certain physical quantities which defines the corresponding critical exponents \cite{Wett_exp}.
In order to get the critical exponent $\nu$ of the correlation length $\xi$
one usually should solve the stationary RG equations around some crossover fixed points \cite{exp}.
In the d-dimensional $O(N)$ model the Wilson-Fisher (WF) fixed point \cite{Nagy_ond},
while in 2d sine-Gordon (SG) model \cite{SG,Nagy_sg,Pangon_sg}
the Kosterlitz-Thouless (KT) fixed point \cite{KT,Kosterlitz}
plays the role of the crossover fixed
point \cite{Nagy_deg}. However there exists several models where the phase space
either does not contain any crossover, saddle point like fixed point.
These models also can possess non-trivial phase structure even including a broken
symmetric phase. In these cases it is extremely difficult to determine the order of the
appearing phase transitions.

Among many examples the massive SG (MSG) model has not any fixed points
due to the trivial scaling of the mass. The model is a simple generalization
of the SG model with an auxiliary mass term \cite{Nagy_schw,Nagy_msg,Nagy_qed2,Nandori_msg}.
The UV scaling of the MSG model predicts two phases with an infinite order phase transition, since
one cannot distinguish the model from the original SG model there.
Otherwise it is well known that the model as the bosonized version of the 2d
quantum electrodynamics (QED$_2$) \cite{boson} possesses a finite order
phase transition. We note that the QED$_2$ is one the most widely used toy
model of self-interacting gauge theories since the model exhibits the soft mechanism of the
confinement \cite{Fishler,Nagy_schw}, and it can be related to the chiral condensate
\cite{Christiansen}. The phases of the bosonized version of QED$_2$ was first mapped with
RG method in \cite{Nagy_msg}, which successfully recovered the phase
structure of the original fermionic model, moreover it was shown that
the critical ratio $\tilde u/\tilde M^2$ in the IR regime exhibits a quantitative
agreement with the value obtained by density matrix RG method \cite{Byrnes}.
The functional RG calculation was performed in the lowest order of the gradient
expansion i.e. in the local potential approximation (LPA).
This result was based on considering the upper harmonics of the periodic term in the
MSG model, furthermore it was assumed that the RG flows run into
singularity \cite{Nagy_msg,Nandori_comp}. However the phase structure can be uncovered
without taking into account the higher harmonics \cite{Nandori_msg},
although the inclusion of wavefunction renormalization disables us to map the scalar model onto
QED$_2$, due to the scaling of the parameter $\beta$ \cite{Nandori_msg}.
The mass term of the MSG model on one hand breaks the periodic
symmetry, and, on the other hand sets the RG evolution of all the couplings relevant below
the mass scale. The broken periodic symmetry brings the KT type transition of the SG model
into a second order one in the MSG model. It is believed \cite{Byrnes} that the model
belongs to the 2d Ising universality class due to its $Z(2)$ symmetry. Here we show that
the exponents does not give the 2d Ising ones suggesting that the model may not belong
to the 2d Ising universality class.

The layered SG (LSG) model is an other generalization of the SG model,
which  contains $N$ scalar fields interacting through a mass matrix \cite{Nandori_uvlsg}.
The model preserves the periodic symmetry of the original model.
The LSG model is the bosonized
version of the multi-flavor QED$_2$, where the number of layers of the LSG model
equals the number of flavors in QED$_2$. The LSG model can also be used to describe
the vortex behaviour of magnetically coupled layered superconductors, e.g.
artificially produced superlattices, where the vortices interact through
the transverse magnetic field \cite{Buzdin}. In this treatment one should
assume that the Josephson coupling due to Cooper pair tunneling between the
superconducting layers is suppressed in the model by the relatively large insulating layers.

The LSG model also exhibits two phases as was demonstrated by using the functional RG method
in an extended UV regime \cite{Nandori_uvlsg}. It was also shown that
the phases are separated by a critical parameter $\beta_c$ which depends similarly on the
layer number $N$ as the critical temperature of
fractional-flux in magnetically coupled high-temperature superconductors.
The critical parameter separates the symmetric and the broken symmetric
phases of the model. It is assumed that there is a KT type
phase transition between the phases \cite{Babaev}, but an LPA investigation cannot
account for that. It was shown in the case of the SG model
\cite{Nagy_zsg} that one should include the next order of the gradient expansion, i.e.
the wavefunction renormalization \cite{mom} in order to unfold the KT-type crossover scaling regime
\cite{Amit,essential,Nagy_zsg}, therefore it seems unavoidable to go beyond LPA
to establish the KT like character of the phase transition in the LSG model.
However, similarly to the MSG model, RG flow equations for the LSG model have no crossover
fixed point because of the trivial evolution of the interlayer coupling.
We note that the due to the inclusion of the wavefunction renormalization the parameter
$\beta$ evolves, which makes impossible to map the LSG model onto the
multi-flavor QED$_2$.

The RG flow equations of the MSG and the LSG models have no crossover fixed points disabling us
to determine $\nu$ in a usual way \cite{exp}. However it was shown in several
scalar model examples \cite{Nagy_deg,Nagy_ond} that the IR scaling of $\xi$
coincides with the one found around the crossover fixed point due to
the global nature of the appearing condensate in the broken symmetric phase \cite{symbreak}.
Our method in \cite{Nagy_deg,Nagy_ond} was capable of describing both the Ising-type
and the KT-type phase transitions and scalings giving us a possibility to find
the exponent $\nu$ in the IR limit. There we define the correlation length as the reciprocal
of that momentum scale $k$ where the evolution stops in the broken symmetric phase due to the 
condensate induced degeneracy. We determine the type of the phase
transition and the corresponding universality class
according to the scaling of $\xi$ and the value of the exponent $\nu$.
We emphasize that this method is the only possibility to determine
the type of the phase transitions for the MSG and the LSG models directly.
Furthermore the usual technique based on characterizing the phase transitions
around a crossover scaling regime is not reliable, since the RG flow should
collect all the quantum fluctuations during the evolution therefore one should
follow the RG flows to the deep IR region. This fact makes the finding of
any infinite order phase transition rather accidental for example in the SG model around
the KT type fixed point. Fortunately the crossover and the IR scaling regimes
provide the same exponent $\nu$ due to the global nature of the condensate
which is closely related to the correlation length in the broken symmetric phase.

For the MSG model we find a second order phase transition characterized by the exponent $\nu=0.5$.
This suggests the possibility of exhibiting a 2d Ising-type phase transition there,
which should give $\nu=1$. We investigate the phase structure of the LSG model in a great detail.
We show that the LPA approximation seems to show two phases in accordance with
\cite{Fishler,Nagy_schw}. However the inclusion of the wavefunction renormalization
changes drastically the IR behaviour of the model, and the broken symmetric phase
melts into a single phase. We also determine the scaling of the correlation length
which proves that the model possesses a KT type phase transition.

The organization of this paper is the following. First the scaling in a finite
order phase transition is investigated in the framework of the
MSG model in \sect{sec:fin}. In \sect{sec:inf} the LSG model is discussed.
Finally, \sect{sec:sum} contains the summary of the results.

\section{Finite order phase transition}\label{sec:fin}

The successive elimination of the quantum fluctuations is performed by means
of the the Wetterich RG equation for the effective action \cite{Wetterich}
\beq\label{WRG}
\dot\Gamma=\hf\mr{Tr}\frac{\dot R}{R+\Gamma''}
\eeq
where $^\prime=\partial/\partial\varphi$, $\dot \ = k\partial_k$ and the trace
Tr denotes the integration over all momenta.
\eqn{WRG} has been solved over the functional subspace defined by the ansatz
\beq\label{eaans}
\Gamma_k = \int_x\left[\frac{Z}2 (\partial_\mu\varphi)^2+V\right],
\eeq
where the potential $V$ has the form
\beq
V = \hf m^2\varphi^2+\sum_{n=1}^{N_u} u_n\cos(n\varphi),
\eeq
for the MSG model, with mass $m=m_k$, the coupling $u_n=u_{n,k}$ and the wavefunction
renormalization $Z=Z_k(\varphi)$. We introduce the higher harmonics of the periodic term
of the potential, since these terms are generated during the RG evolution.
The ansatz for the wavefunction renormalization $Z$ contains a constant
and a periodic term, i.e.
\beq
Z = z_0+\sum_{n=1}^{N_z} z_n\cos(n\varphi).
\eeq
We note that, as opposed to the polynomial models, the field independent
coupling $z_0$ has a non-trivial evolution without considering the upper harmonics $z_n$,
$n=1...N_z$. The polynomial suppression has the form
\beq\label{polsup}
R = p^2\left(\frac{k^2}{p^2}\right)^b,
\eeq
with $b\ge 1$. The choice $b=1$ coincides with the usual Callan-Symanzik scheme \cite{CS}.
From the functional RG equation in \eqn{WRG} one can deduce evolution equations for the
couplings $m^2$, $u_n$ and $z_n$. For shorthand we introduce
$u\equiv u_1$ and $z\equiv z_0$. The evolution of mass decouples from the
other RG flow equations, and one obtains $m^2 = m_\Lambda^2$, where $\Lambda$
is the UV cutoff. The further couplings scale according to the equations
\bea\label{eecoup}
\dot u_n &=& \hf{\cal P}_n \int_p \dot R G\nn
\dot z_n &=& \hf{\cal P}_n \int_p \dot R \biggl[
-Z''G^2+\left(\frac2{d}Z'^2p^2+4Z'\Gamma'''\right)G^3\nn
&&-2\left[\Gamma'''^2\left(\partial_{p^2}P+\frac2{d}p^2\partial^2 P\right)
+\frac4{d}Z'p^2\Gamma'''\partial_{p^2}P\right]G^4\nn
&&+\frac8{d}p^2\Gamma'''^2\partial_{p^2}P^2G^5\biggr]
\eea
with $G=1/(Zp^2+R+V'')$ and $P = Zp^2+R$. We introduced the projection as
${\cal P}_n=\int_{\varphi}\cos(n\varphi)/{\pi}$ for $n> 1$ and
${\cal P}_0=\int_{\varphi}/{2\pi}$.
The momentum and the projection integrals have to be performed numerically.
The scale $k$ covers the momentum interval from the UV cutoff $\Lambda$ to zero. 
Typically we set $\Lambda=1$.

The RG evolution becomes singular, i.e. the r.h.s. of the RG flow equation is infinite in
\eqn{eecoup} when
\beq\label{degen}
Zp^2+R+V'' = 0.
\eeq
The non-trivial expectation value emerges from the average $\langle\varphi\rangle=0$ continuously,
therefore one can set $\varphi=0$, giving
\beq\label{degen2}
\bar k^2 - \sum_{n=1}^{N_u}n^2 u_n=0,
\eeq
where $\bar k =\min P$ has been introduced. By using the polynomial suppression in \eqn{polsup}
one arrives at
\beq
\bar k^2 = b k^2\left(\frac{\sum_{n=0}^{N_z}z_n}{b-1}\right)^{1-1/b},
\eeq
for $b=1$ we have $\bar k = k$. We introduce furthermore the normalized fundamental mode
$\bar u_n = u_n/\bar k^2$.

\subsection{Infrared scaling}\label{sec:WFR}

We numerically calculated the flows of the coupling $\bar u$ in the presence of power-like
suppression scheme. The mass scales trivially, i.e. $\tilde m\sim k^{-1}$. We considered
$N_u=5$ and $N_z=1$ for the number of couplings.

The coupling scales as $\tilde u\sim k^{-2}$ below the mass scale
in the IR region, independently on any initial UV microscopic parameters.
However the model has two phases \cite{Nagy_msg,Nandori_msg,Byrnes},
separated by the critical ratio $u_{0c}/m^2_0$, where $u_{0c}$
is the critical IR limiting value of the coupling $u$, which separates
the appearing two phases in the model. Since the mass $m$ does not evolve,
then $m=m_0$. The phases can be distinguished
by considering the sensitivity of the effective potential to the microscopic
parameter $u_{\Lambda}$ \cite{SG,Nagy_msg,Nagy_qed2,Nagy_zsg}. In the symmetric
phase the effective potential depends on the UV value $u_{0\Lambda}$, while
in the broken symmetric phase the effective potential is is insensitive to
any parameters, i.e. it is (super)universal.

\begin{figure}[ht] 
\begin{center} 
\epsfig{file=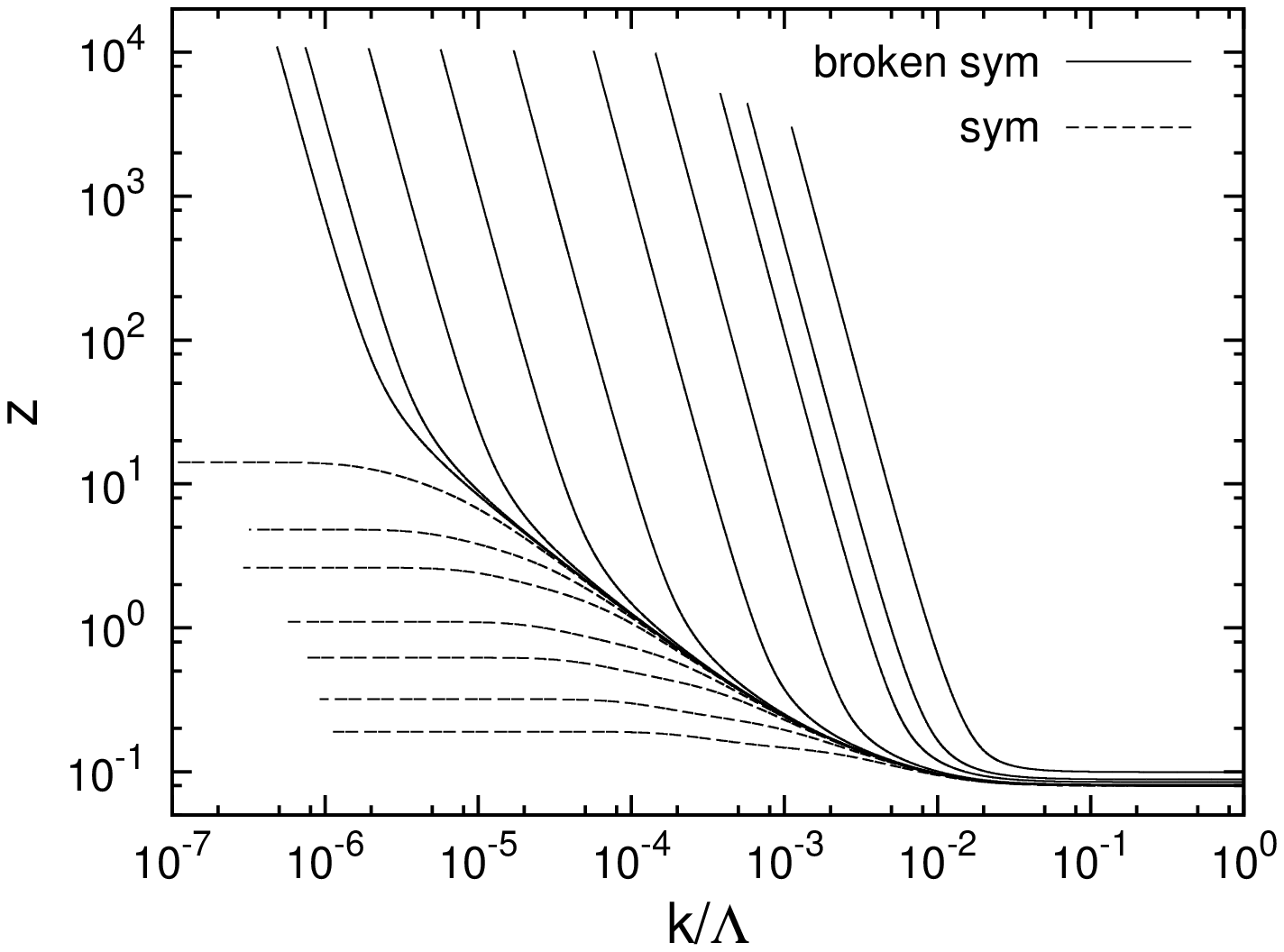,width=6cm,angle=-90}\\
\epsfig{file=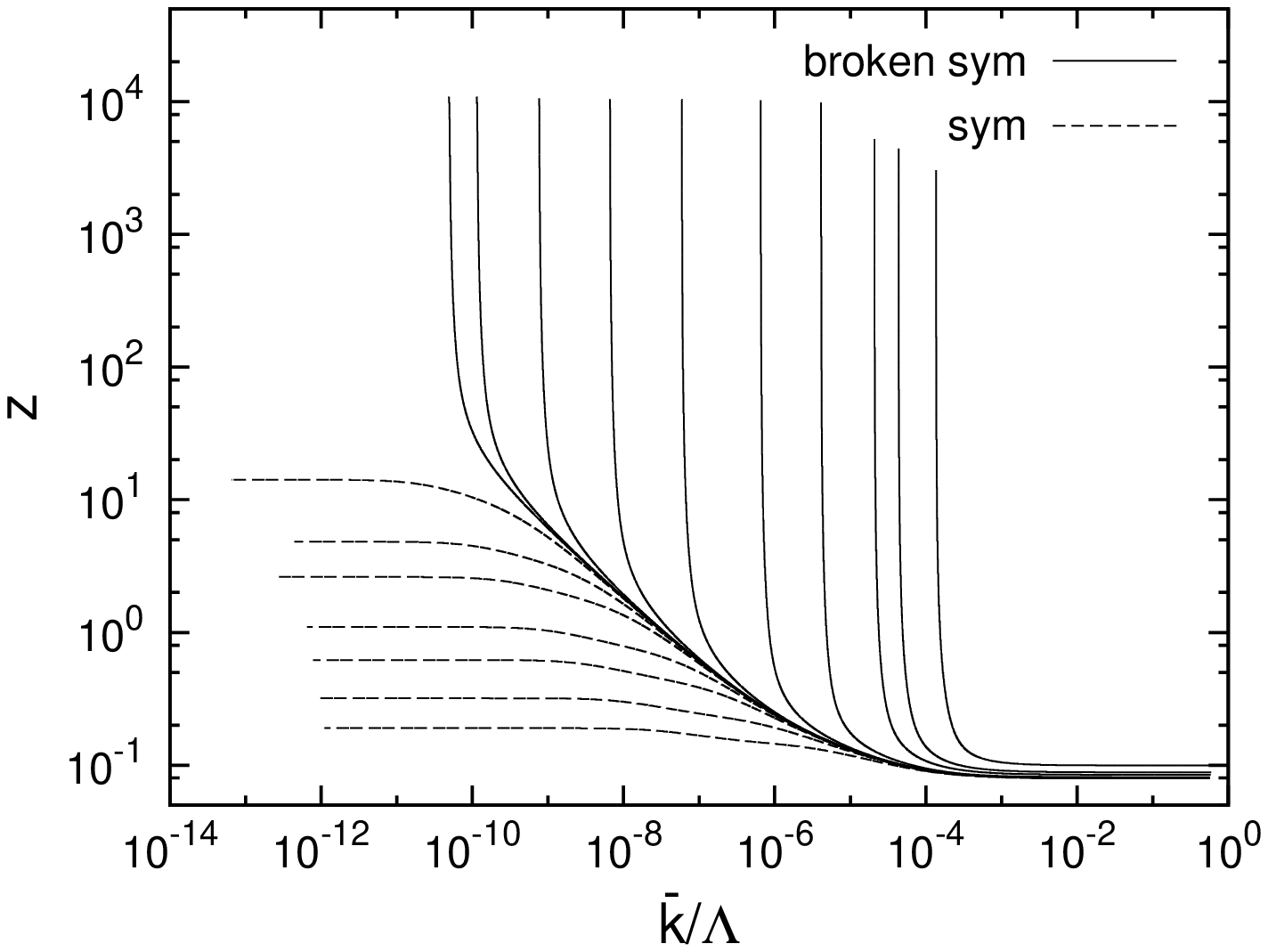,width=6cm,angle=-90}
\caption{\label{fig:zmsgz}
The scaling of the wavefunction renormalization $z$ according to the two scales $k$ (up) and
$\bar k$ (down) for the choice $b=2$.
}
\end{center}
\end{figure}

\begin{figure}[ht] 
\begin{center} 
\epsfig{file=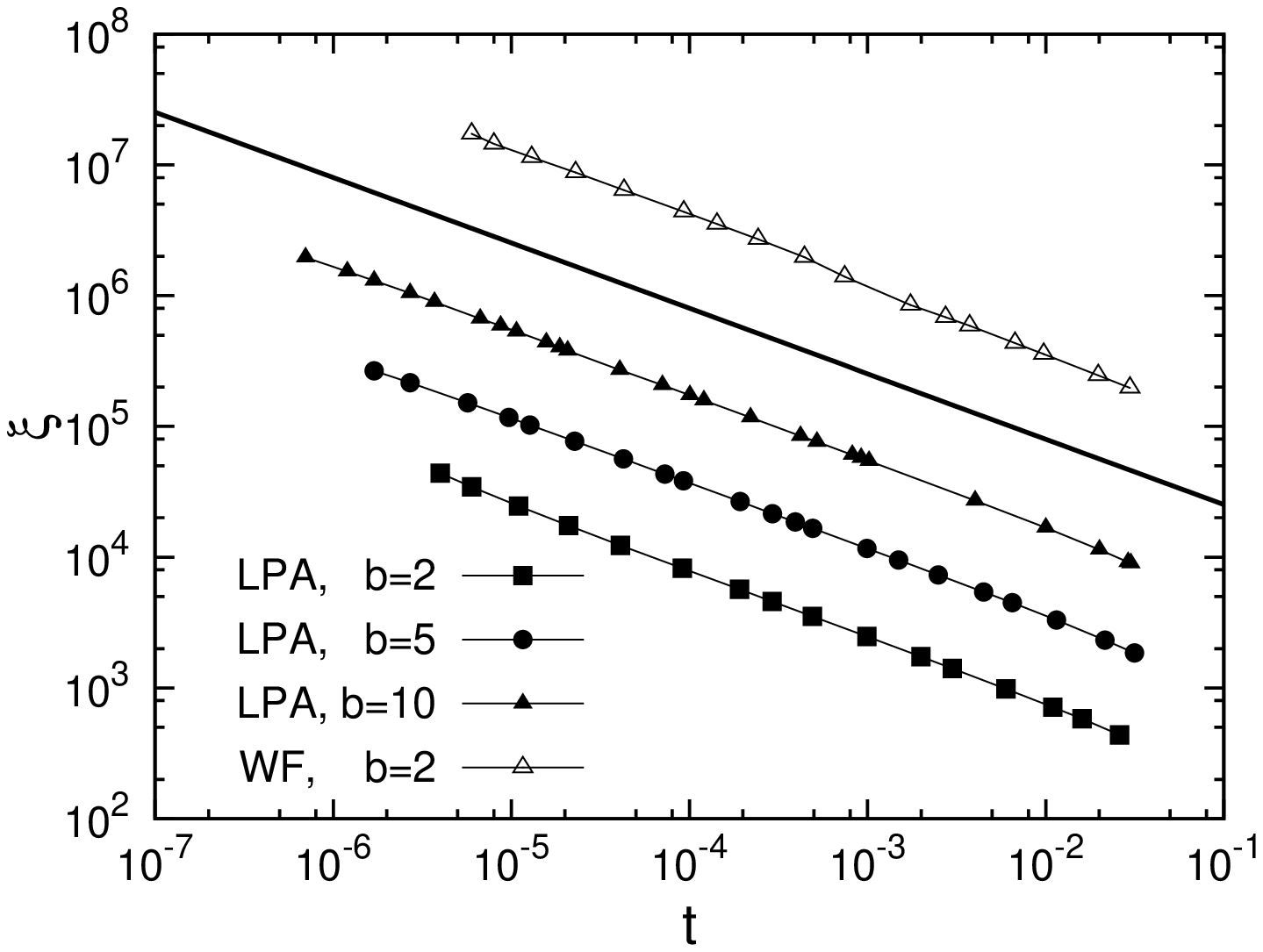,width=6cm,angle=-90}
\caption{\label{fig:zmsgexp2}
The reduced temperature dependence of the correlation length for different suppression
schemes, WF refers to the numerical solution including the wavefunction renormalization.
}
\end{center}
\end{figure}

Setting the value of the mass $m$ fixed one has a broken symmetric phase when
$\tilde u_{\Lambda}>\tilde u_{\Lambda~c}$. \fig{fig:zmsgz} shows that
one can also distinguish the two phases of the model according to the scaling of $z$.
In the symmetric phase $z$ is constant in the UV, scales as $z\sim k^{-\eta_{co}}$
in the crossover and is constant too in the IR limit. The crossover scaling appears
below the mass scale far from the deep IR regime. 
The anomalous dimension $\eta$ being characteristic for the divergence of the correlation
function of the field variables, defined as  $\eta=-d\log z/d\log k$ is not unique
during the evolution, and its value is sensitive to the parameter choice of $b$, too.
The final constant value of $z$ in the IR limit diverges as the initial parameters
approach the critical ratio. In the broken symmetric phase we can find all the three regimes
for $z$ as in the symmetric case, but the IR regime shows its relevant scaling with a high value
of anomalous dimension $\eta_{IR}$. We also plotted the evolution of $z$ as the function $\bar k$.
In the symmetric phase the flow is qualitatively same that was obtained for $k$.
However the crossover scaling gives universal, i.e. $b$ independent exponent
$\bar\eta_{co}=1$ and in the broken symmetric phase the flows blow up in the IR, giving
$\bar\eta_{IR}=\infty$. This defines a dynamical momentum scale $\bar k_c$, where
the RG evolution stops and the effective potential becomes degenerate. We got similar
behaviour for other scalar models, too \cite{Nagy_deg,Nagy_ond}. The reciprocal
of $\bar k_c$ is identified as the correlation length, i.e. $\xi =1/\bar k_c$.
The initial UV value of $z_{0\Lambda}$ can be identified with the temperature,
therefore its distance from its critical value (for a fixed value of $u_{\Lambda}$)
provides the reduced temperature $t$.
This makes us the possibility to determine the $t$ dependence of the correlation length.
In \fig{fig:zmsgexp2} one can see how $\xi$ depends on the  critical temperature $t$.
We performed the calculation for $N_u=5$ and $N_z=1$ and obtained that
the correlation length scales as a power-law function giving straight lines in
log-log scale according to
\beq
\xi\sim t^{-\nu}
\eeq
with $\nu\approx 1/2$. We repeated our calculations for the case $N_u=1$ and $N_z=0$ and
and for several values of $b$, and we got the same value for the exponent.
This means that neither the upper harmonics nor
the field dependent wavefunction renormalization $Z$ affects the value of $\nu$.
This result is in accordance with the ones obtained for the SG model where
we also got such sensitiveness of $\nu$ to the upper harmonics.
We also made calculation in the case of LPA, which also gave $\nu=1/2$.

Our results seems to contradict with results obtained for the 2d Ising exponent,
where $\nu=1$. In \cite{Byrnes} it was shown that QED$_2$ is in the 2d Ising universality
class, and it is argued that the MSG model has a $Z(2)$ symmetry which implies that the
model is in the 2d Ising universality class, too.
Although the MSG model has only a $Z(2)$
symmetry, the RG flow equations for the couplings $u_n$ seems to preserve the periodicity.
This may affect the simple $Z(2)$ symmetry and may eventually
give that the MSG model is not in the 2d Ising the universality class.

These results demonstrate that one can get a clear picture from the type of the
phase transition in the MSG model according to its IR scaling. Since the model
has no crossover fixed point this method is the only possibility to uncover the order
of the phase transition there.

Considering the RG evolution equations in \eqn{eecoup} one can easily see that
we have no ``static'' solution, which means that in mathematical sense we have
no nontrivial fixed points in the MSG model. So far we investigated such models where the
IR value of $\nu$ was inherited from its crossover value \cite{Nagy_deg,Nagy_ond}.
In the MSG model it is not the case. This fact clearly shows the power
of finding the scaling relations in the IR.
We conclude that the IR scaling gives us an exclusive and correct picture for the
scaling of the correlation length.

\section{Infinite order phase transition}\label{sec:inf}

The LSG model can be described by the the corresponding effective action \cite{layer}
\beq\label{effac}
\Gamma_k = \int_x\left[\frac{z}2\sum_{n=1}^N(\partial_\mu\varphi_n)^2+V\right],
\eeq
with the scalar fields of the layers $\varphi_n$, $n=1\ldots$, the
wavefunction renormalization $z$, and the potential $V$ of the form
\beq\label{Vans}
V = \hf J \left(\sum_{n=1}^N\varphi_n\right)^2+u\sum_{n=1}^N\cos\varphi_n,
\eeq
with $J$ the interlayer coupling and a single coupling $u$ for every layer.
The elementary excitations with higher vorticity, which are coupled to the operators like
$\cos(n\varphi_m)$, with $n>1$ and $m=1\ldots N$ are negligible in the determination
of the thermodynamic properties of the model. Otherwise, the fundamental mode $u$ is capable of
mapping the phase structure of the model \cite{Nandori_uvlsg,Nagy_qed2,Kovacs}.
We consider only a field independent wavefunction renormalization constant $z$, the
field dependent ones do not affect the results in the case of periodic models.

In order to find the flows we use the same Wetterich equation in \eqn{WRG} with the matrix
\beq
\Gamma''=zp^2\dblone+V''
\eeq
and $V''=\delta^2 V_k/\delta\varphi_i\delta\varphi_j$, $i,j=1\ldots N$
and now the trace Tr denotes the summation over the internal index
(over the components of the scalar field) and the integration over all momenta.
Now we set $b=1$ in the suppression if the wavefunction renormalization $z$
also evolves, since although it gives moderate convergence and
needs high computational accuracy, it makes the evolution equations much simpler.
The RG evolution equation in \eqn{WRG} gives differential equations for the
dimensionful couplings $z,u$ and $J$. The latter evolves trivially i.e.
\beq\label{evolJ}
\dot J = 0.
\eeq
For the potential we have
\beq\label{evolV}
\dot V_k = \hf \Tr[\dot R G]=\hf\tr\int_p\frac{\dot R}{ zp^2\dblone+R\dblone+V''}
\eeq
where $\tr$ is a trace over the components of the scalar field and $\dblone$ denotes
the $N$th order unit matrix. The evolution of $u$
can be picked up by the acting on \eqn{evolV} with the projection operator
\beq
{\cal P}_N = \frac2{(2\pi)^N}\int_{\varphi_1\ldots\varphi_N}
\cos\varphi_2\ldots\cos\varphi_N,
\eeq
to get the corresponding Fourier mode. The evolution of $z$ becomes
\bea\label{evolz}
\dot z &=& \sum_{n=1}^N\frac2{N}\int_p k^2 \tr\left[G_p V''^{(n)}G_pV''^{(n)}\left(2p^2z^2
G_p^3{-}zG_p^2\right)\right].\nn
\eea
for the choice $b=1$, where $V''^{(n)}=\delta V''/\delta\varphi_n$.

\subsection{Local potential approximation}

We solved the RG evolution equation in Eqs. (\ref{evolJ}), (\ref{evolV}) and (\ref{evolz})
for the LSG model  numerically. We determined
the RG flows by taking into account only a single mode in the potential, but without taking any
further approximations. The calculation is carried out for bi-layered systems, i.e. $N=2$.
This specification on one hand is necessary because the momentum
integral and the integrals in the projection ${\cal P}_N$ has to be performed with no
truncation in $u$ in order to get the deep IR evolution. On the other hand the qualitative
results are not expected to change by increasing the number of layers.

We are interested in the LPA results because the bosonization technique makes
the LSG model equivalent to the multi-flavor QED$_2$. It is assumed that
the latter model has two phases. The mapping between the models can be performed by
the parameter choice $z=1/4\pi$. This requires the freezing of the wavefunction
renormalization and confines our treatment to LPA approximation \cite{Kovacs}.

Acting on the projection ${\cal P}_2$ to the evolution equation for the potential
in \eqn{evolV}, one obtains
\beq
\dot u = \int_{p,\varphi_2\varphi_2}\frac{\dot R(2P-u(\cos\phi_1+\cos\phi_2))\cos\varphi_2}
{(P-u\cos\varphi_1)(P-u\cos\varphi_2)-J^2}
\eeq
with $P=zp^2+R+J$. The interlayer coupling $J$ does not evolve, $J=J_\Lambda$.
In the LPA treatment the evolution of $z$ is also frozen.
The extended UV scaling gives \cite{Nandori_uvlsg,Kovacs}
\beq\label{qed2yUV}
u = u_\Lambda
\left(\frac{k}{\Lambda}\right)^{1/8\pi z}
\left(\frac{k^2 + 2 J}{\Lambda^2 + 2 J}\right)^{1/16\pi z}.
\eeq
The coupling $\tilde u=u/k^2$ scales irrelevantly (relevantly) in the symmetric (broken symmetric)
phase if $1/z>16\pi$ ($1/z<16\pi$), respectively. We note, that generally, the critical
frequency which separates the two phases of the model is \cite{Nandori_uvlsg}
\beq
\label{laydepz}
z_c = \frac{N-1}{8\pi N}.
\eeq
In the broken symmetric phase the RG flow of the couplings can run into singularity if the
numerical calculation is not precise enough. For example in the case of the SG model the
exact treatment in the coupling $u$ is necessary to avoid the singularity for certain
suppression schemes \cite{Nagy_zsg,Nandori_uvlsg}. Therefore we should include all the
contributions of the coupling $u$ exactly for the LSG model, too.

Similarly to the MSG model we introduce $\bar k =\min P$, the dimensionless couplings
$\tilde u = u/k^2$ and $\bar u = u/\bar k^2$. The latter tends to 1 if the flow approaches
the degeneracy.

\subsubsection{The phase structure}

\begin{figure}[ht] 
\begin{center} 
\epsfig{file=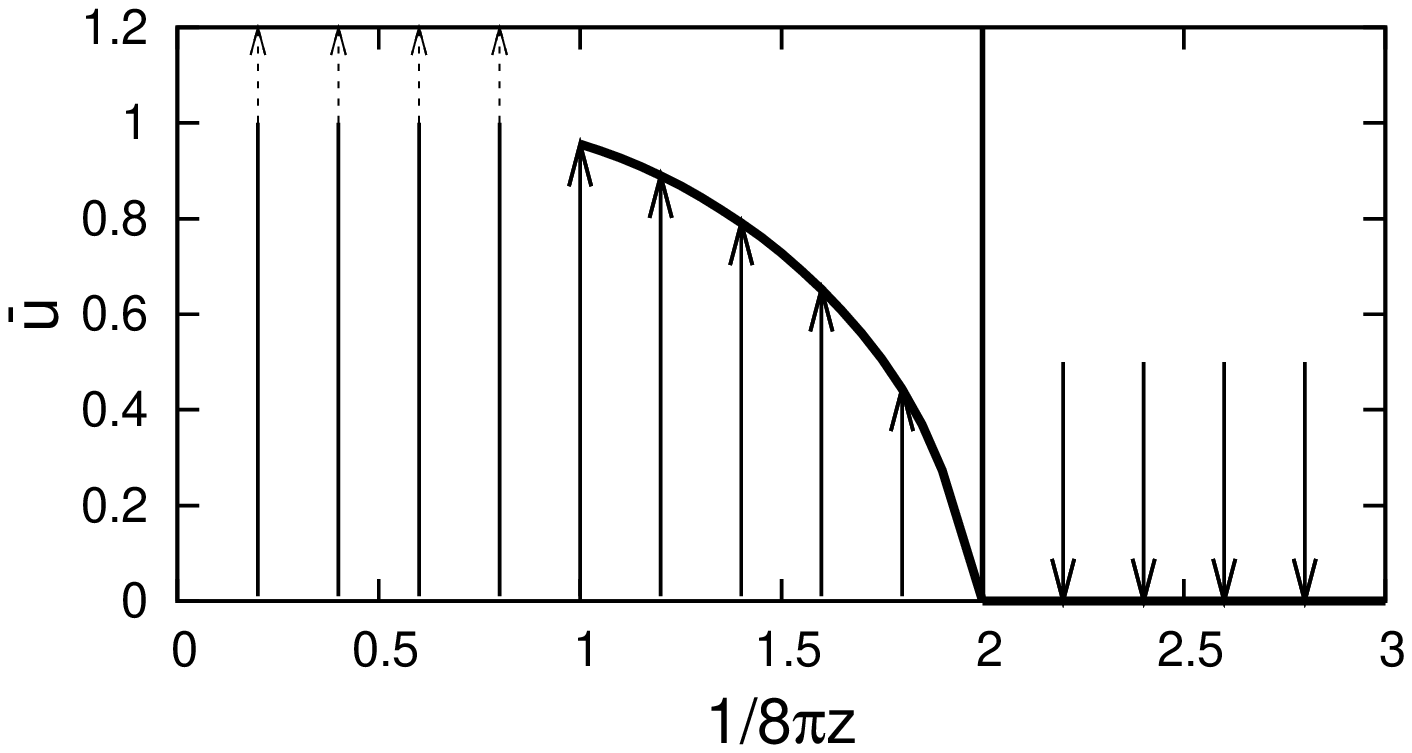,width=4.6cm,angle=-90}
\caption{\label{fig:lpaphase}
The phase structure of the 2-layer SG model is shown in LPA, $b=2$. The vertical line shows
the separatrix at $z=1/16\pi$. The scaling of the coupling $\bar u$ is relevant (irrelevant)
in the broken symmetric (symmetric) phase, respectively. When $z>1/8\pi$ we have no information
about the deep IR scaling.}
\end{center}
\end{figure}

The extended UV results suggest that the model has two phases. In \fig{fig:lpaphase} we
plotted the trajectories of the phase space at a certain value of interlayer coupling $J=10^{-4}$.
We should admit that the phase space is 3-dimensional, but $\bar J$ scales trivially, therefore
we always project the phase space to the other axes.
There is a symmetric phase at $z<1/16\pi$. The flow of the coupling $\bar u$ is irrelevant, i.e.
it tends to zero in the deep IR regime. Although the flows should span the momentum
interval $k\in[\Lambda,0]$ the evolution cannot reach its lower limit, because we should keep
$\Delta k/k$ small throughout the evolution.
We note however that only the deep IR scaling of the couplings which provides us all
the information about the effective potential of the model. The numerical results
show that in the
deep IR the coupling is sensitive to its initial value $\bar u_\Lambda$. There is a line of
fixed points in the symmetric phase at $z<1/16\pi$ and $\bar u=0$, which is
denoted by thick straight line in \fig{fig:lpaphase}.

\begin{figure}[ht] 
\begin{center} 
\epsfig{file=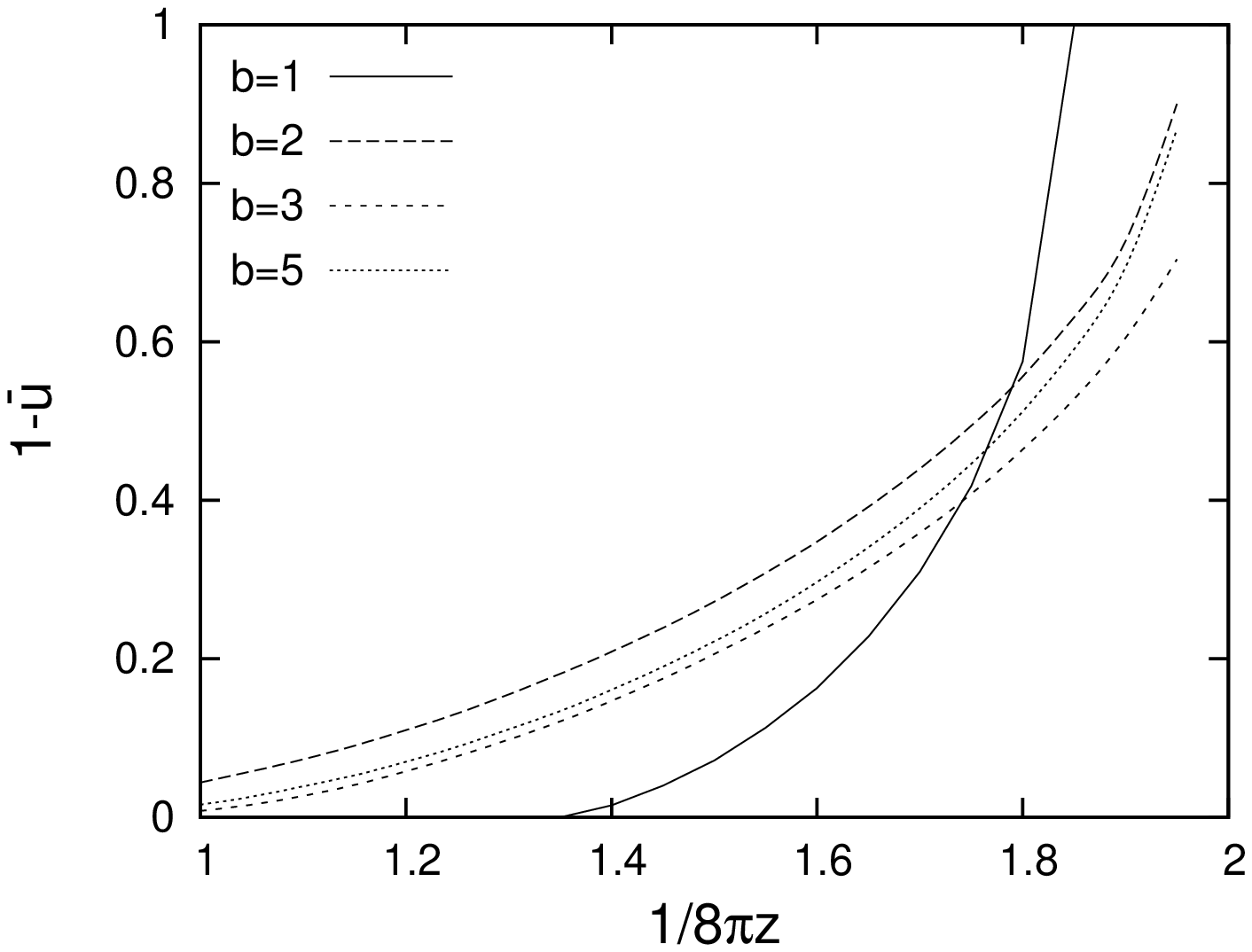,width=6cm,angle=-90}
\caption{\label{fig:lpascheme}
The scheme dependence of the IR fixed point in the case of  $1/8\pi>z>1/16\pi$
is presented. The choice $b=1$ gives one of the worst, but fastest evolution.
The different schemes give qualitatively same phase structure.}
\end{center}
\end{figure}

\begin{figure}[ht] 
\begin{center} 
\epsfig{file=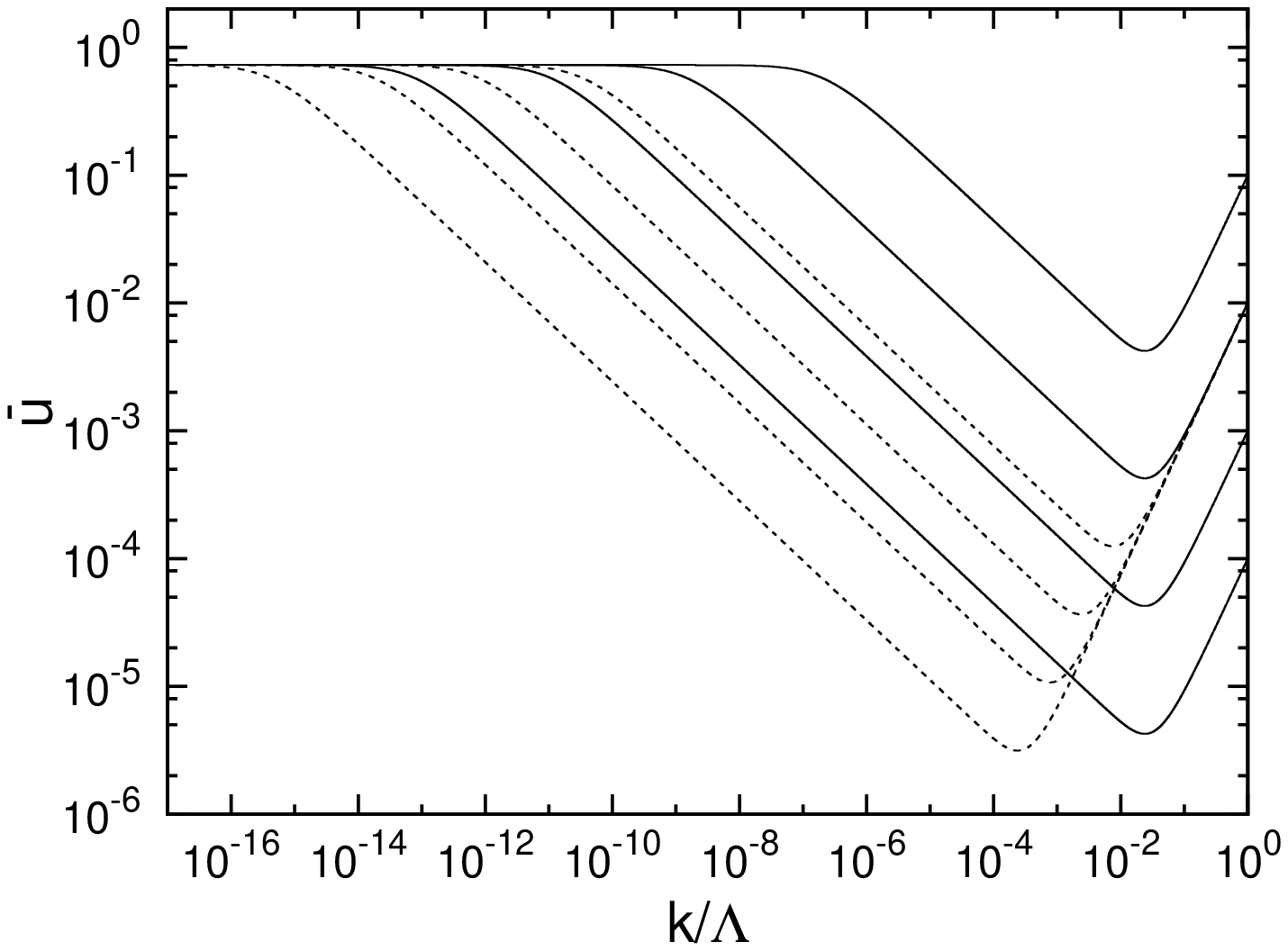,width=6cm,angle=-90}
\caption{\label{fig:lpabrokenhigh}
The evolution of the coupling $\bar u$, with the parameter choice $z=1/12\pi$ and $b=2$.
The flows go to the same IR fixed point independently on the initial value of
$\bar u_\Lambda=$ and $J$.}
\end{center}
\end{figure}

\begin{figure}[ht] 
\begin{center} 
\epsfig{file=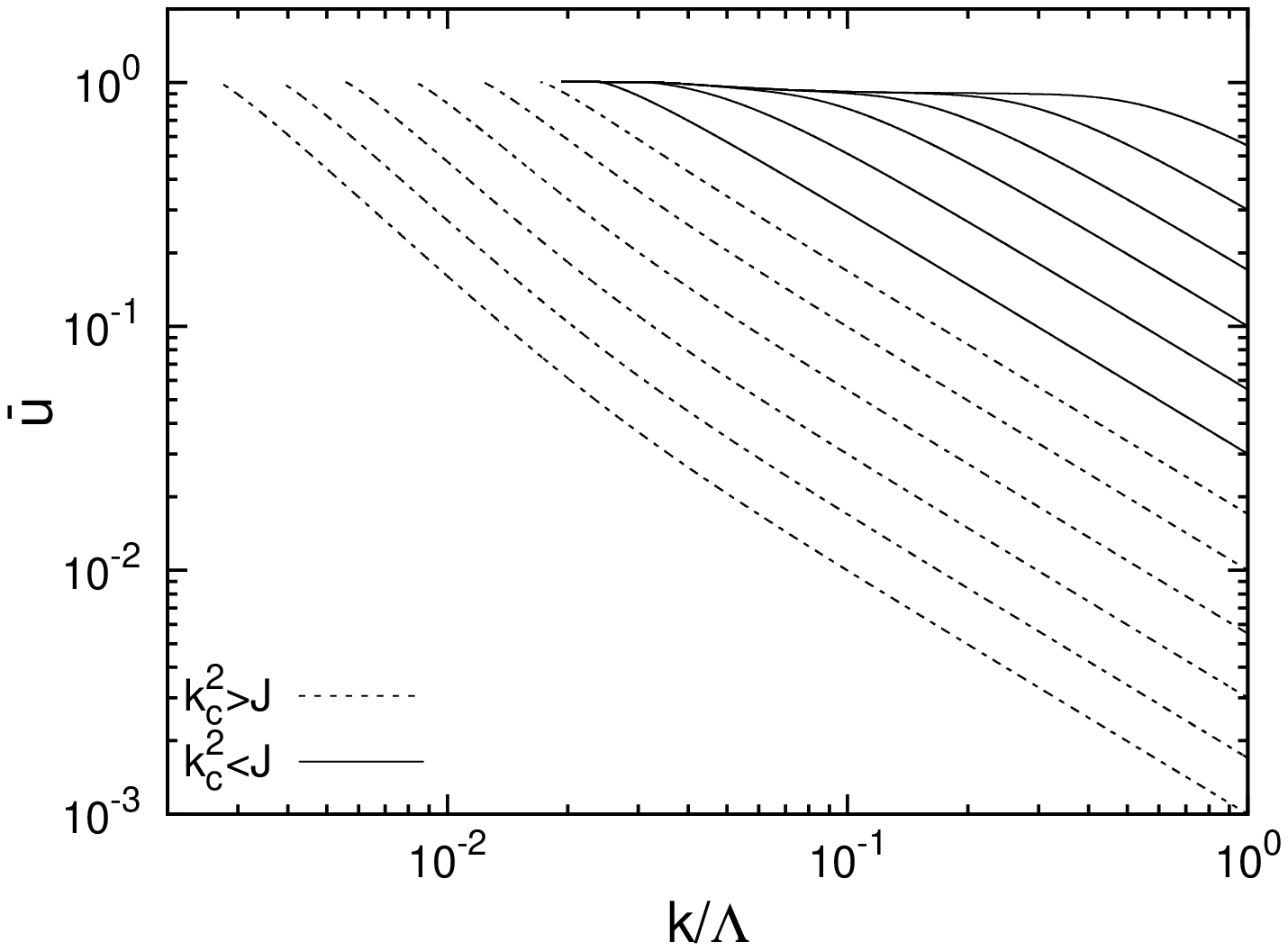,width=6cm,angle=-90}
\caption{\label{fig:lpabrokenlow}
The evolution of the coupling $\bar u$, with the parameter choice $z=1/4\pi$ and $b=2$.
The flows run into singularity, independently on the on the initial value of
$\bar u_\Lambda=$ and $J$.}
\end{center}
\end{figure}

When $z>1/16\pi$, then the RG flow of the coupling $\bar u$ belongs to the
the broken symmetric phase. \fig{fig:lpaphase} shows that if
$1/8\pi>z>1/16\pi$ then $\bar u$ tends to an IR fixed point denoted by a thick solid line
there. We note that the concrete value of these fixed point is scheme dependent,
as is demonstrated in \fig{fig:lpascheme}.
This behaviour is quite similar to that was obtained in the case of the SG model \cite{Nagy_deg}.
The singularity is avoidable by an appropriate choice of the parameter $b$. The IR fixed
points are farest from the singularity if $b=2$. In this sense this gives the most stable
evolution and an optimized value for $b$. Let us note, that the optimized value is usually
$b=2$ \cite{Nandori_msg,Litim}.
The evolution of $\bar u$ starts with an irrelevant scaling due to the SG nature of the model
in the UV \cite{Nandori_uvlsg}. In \fig{fig:lpabrokenhigh} one can follow that the evolution
turns to relevant at about the order of the mass scale $\sqrt{2J}$.
After the relevant scaling regime in the crossover the coupling flows to a fixed point
marginally. There is a certain scale $k_c$, where the marginal scaling appears. This scale
can be identified with the appearance of the global condensate in the model. The value of
the IR fixed point depends on $z$ and with line of fixed points in he symmetric
phase they constitute a continuous line starting from $z=1/8\pi$, $\bar u=1$ to $z=1/16\pi$,
$\bar u=0$ as showed with a thick curve in \fig{fig:lpaphase}. The weak point of the LPA
investigation can be found in the region $z>1/8\pi$. \fig{fig:lpabrokenlow}
shows some typical evolutions there.
Unfortunately the evolution always stops since the coupling reaches the singularity condition
in \eqn{degen2}. If $k_c^2<J$ (dotted lines in \fig{fig:lpabrokenlow}) then the coupling scales
as $\bar u\sim k^2$ below the scale $\sqrt{2J}$ therefore the evolution runs into singularity.
If $k_c^2>J$ the the evolution first scales marginally giving universal crossover value
(see the solid lines in \fig{fig:lpabrokenlow}), but as it reaches the momentum scale
$\sqrt{2J}$ the evolution turns to relevant giving the flow into singularity again.
We note that these scaling laws are independent on the scheme, implying that the deep
IR regime cannot be mapped when $z>1/8\pi$. The results may suggest the splitting of
the broken symmetric regime to two different phases as in the massive SG model
\cite{Nagy_msg,Nandori_msg} and as it was obtained from UV scaling considerations
for the LSG model \cite{Smilga}. However this question cannot be answered
in the framework of the LPA, since the evolutions always run into singularity in there.

Does the RG method fail to map out the phase structure of the LSG model? According to
\fig{fig:lpaphase} the answer is affirmative since we have no information on the deep
IR scaling when $z>1/8\pi$. According to fundamental considerations the RG flows cannot
run into singularity if the effective action is not truncated. However we neglected
the contributions of the higher harmonics, and the gradient expansion in this treatment.
The upper harmonics can account for the physics of the excitations of higher vorticity,
but these excitations are usually negligible and we do not expect them to avoid
the singularity, as in the case of the SG model \cite{Nagy_zsg,Nandori_comp}.
Furthermore the inclusion of the upper harmonics gives only qualitative
changes in the flow. Therefore one should take into account the next term in the gradient
expansion, i.e. the wavefunction renormalization.

\subsection{KT transition}

\begin{figure}[ht] 
\begin{center} 
\epsfig{file=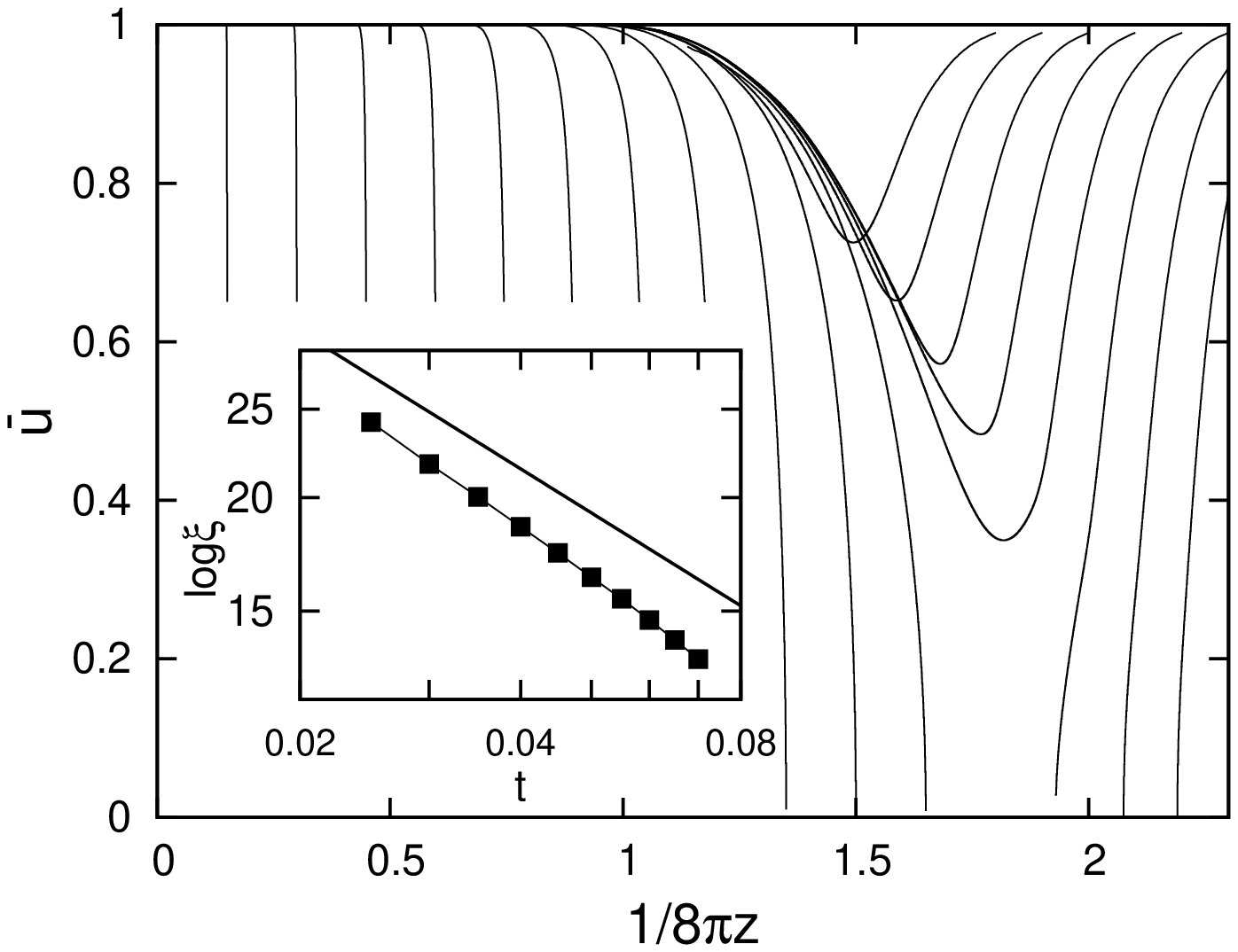,width=6cm,angle=-90}
\caption{\label{fig:zphase}
The phase structure of the 2-layer SG model is shown. The trajectories intersect since
the dimensionless parameter $\bar J$ also evolves.
In the inset the scaling of the correlation length gives $\nu\approx0.57$, the straight
line with the slope $-1/2$ is drawn to guide the eye.}
\end{center}
\end{figure}

Going beyond the LPA and letting $z$ evolve one obtains the typical phase structure of the 
bi-layer SG model plotted in \fig{fig:zphase} for a certain value of $J_\Lambda=0.1$.
The calculations for wavefunction renormalization
were performed with the choice $b=1$. Every other parameter value
would gave qualitatively similar figures. As in the LPA, there is a symmetric phase above
$1/z_\lambda=16\pi$, and a broken symmetric one below this critical value. The symmetric
phase contains a line of IR fixed points similarly to the LPA results. It implies that
the evolution of $z$ is practically negligible in the symmetric phase. Furthermore the
sensitivity of the effective potential to the UV initial values of $\bar u$ also holds.

\begin{figure}[ht] 
\begin{center} 
\epsfig{file=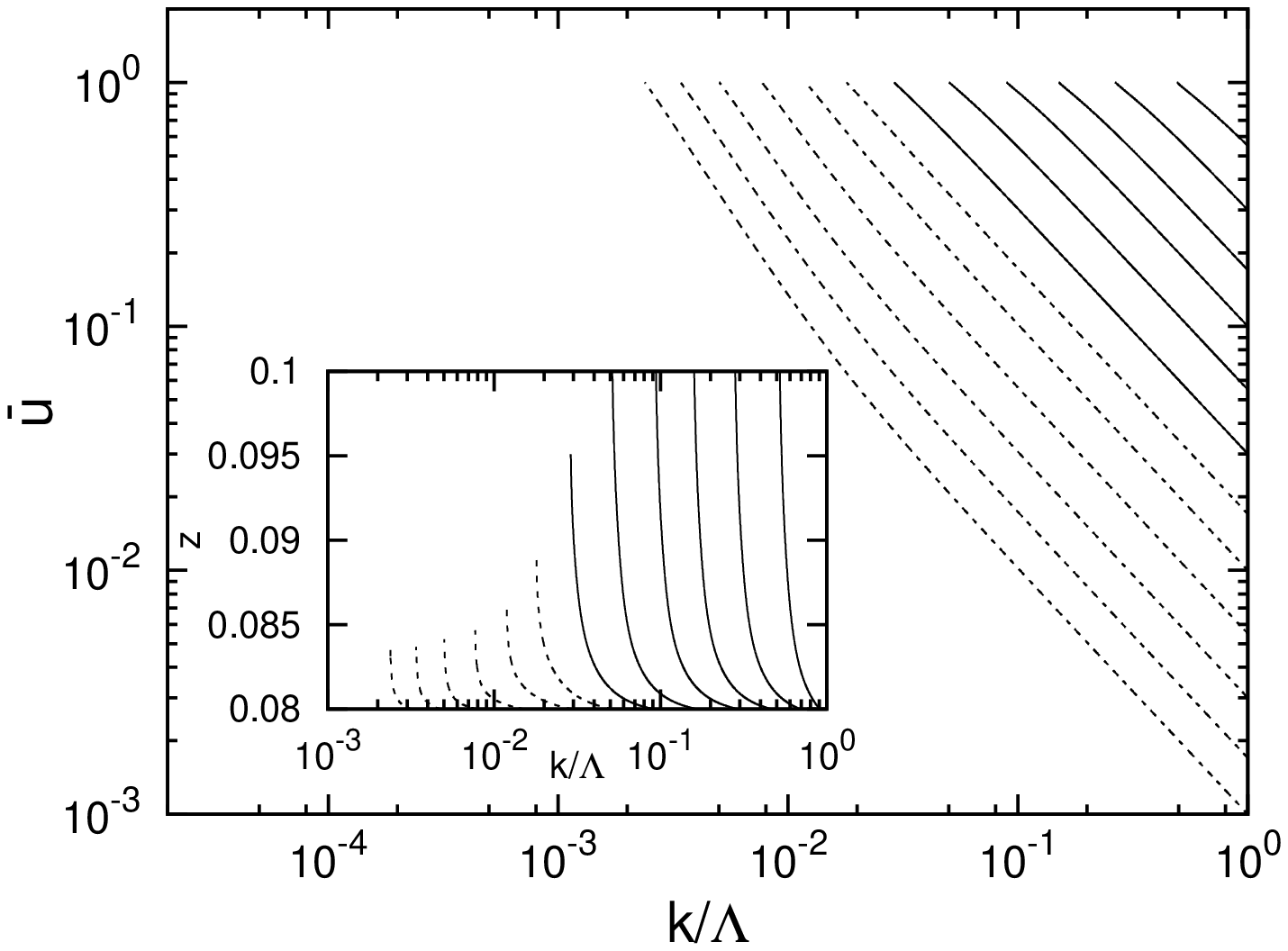,width=6cm,angle=-90}
\caption{\label{fig:zbrokenlow}
The evolution of the coupling $\bar u$ with different initial values of $u\Lambda$ satisfying
$k_c^2>J$ ($k_c^2>J$) and plotted with solid (dashed) line, respectively. In the
inset the evolution $z$ is drawn. At a certain value of $k$ the flows blow up.}
\end{center}
\end{figure}

The broken symmetric phase drastically changes in comparison with LPA. From \fig{fig:zphase} it is
clear that there is an attractive IR fixed point of the model situated at $\bar u=1,1/z=0$ and
the trajectories tend there. It makes no sense to split the broken symmetric phase according to
the phase space in \fig{fig:zphase} as was suggested by the LPA results. Furthermore the figure
shows that $z\to\infty$ in the deep IR regime, which enables to the flow avoid the singularity.
In \fig{fig:zbrokenlow} we plotted the flows of the coupling $\bar u$ starting
from different UV initial values and corresponding to both relations $k_c^2>J$ and $k_c^2>J$.
It seems that the evolutions run into singularity, however they does not. The inset shows that
the evolution of $z$ is practically constant and in the vicinity of the singularity they blow up.
On one hand it brings the IR value of $1/z\to 0$, and on the other hand it stops the evolution
of the coupling at $\bar u=1$. This gives the single IR fixed point in the broken symmetric
phase, showing that opposite to the suggestion of the LPA results, the broken symmetric phase
is unique. This gives a very serious limitation of the usage of LPA in this model. The IR fixed
point satisfies the degeneracy condition in \eqn{degen2}. The appearing effective potential
in superuniversal, i.e. independent of any UV initial parameters. This is a typical property of
the spontaneously broken symmetric phase.

The exponent $\eta$ deduced from the evolution of $z$ is approximately zero in the crossover
regime similarly to the SG model. However approaching the IR fixed point the evolution of
$z$ blows up at a certain scale $\bar k_c$ giving infinitely large $\eta$. Again,
the scale $\bar k_c$ is identified with the reciprocal of the correlation length $\xi$. Its 
dependence on the reduced temperature $t$ can then be easily read off.
In case of infinite order phase transition the correlation length scales as
\beq
\log\xi\sim t^{-\nu}.
\eeq
In the inset of \fig{fig:zphase} we plotted the $t$ dependence of $\log\xi$, and
its slope gives $\nu\approx 0.57$ for the exponent. This proves the infinite nature of the phase
transition of the LSG model. Let us emphasize, that although a KT-type phase transition
is expected in the LSG model, it has not been proven before.

\section{Summary}\label{sec:sum}

We developed a new strategy in order to identify the type of the phase transitions
in field theoretical models. It is based on the fact that there is an IR fixed point
in the broken symmetric phase and the appearing degeneracy during the RG flow defines a dynamical
momentum scale which reciprocal can be identified with the correlation length $\xi$, therefore
the critical exponent $\nu$ of $\xi$ can be determined in an extremely easy way.
The value of $\nu$ may signal the corresponding universality class of the model.
Our new strategy enables us to calculate $\nu$ in such situations
where the RG flow equations has no fixed point, too.

In this article we showed two non-trivial models to demonstrate the power of our new
method. First we investigated the MSG model, where it is known that the inclusion of the
mass term in the action changes the phase structure considerably
as compared to the original, massless one.
Our new method showed that the model possesses a second order but not an Ising-type
phase transition, since we obtained $\nu=1/2$ for the exponent of $\xi$.
As a second example we treated the LSG model. The degeneracy induced scaling strategy
enabled us to show that the model really has a KT-type phase transition as was conjectured
earlier. It was also shown that its broken symmetric phase is unique.

The results of our proposed new method may confirm the quantum censorship conjecture
\cite{Pangon_sg,Polonyi_qc}. A proper choice of the suppression and the inclusion
of the wavefunction renormalization avoids the singularity of the RG flows. The scaling of
the couplings as the function of $\bar k$ stops in the
IR and gives a characteristic length to the appearing degenerate vacuum, or condensate,
while the original momentum scale $k$ reaches zero.

The developed strategy is capable of describing first order phase transitions, too.
In this case the correlation length does not diverge in the broken symmetric phase
if we approach the transition point. Thus the type of the phase transition can be
identified in an easy way according to the scaling around the IR fixed point as it was
demonstrated in the case of quantum Einstein gravity \cite{Nagy_qg}. We note that in this
model the existence of the IR fixed point has been widely investigated recently \cite{ir_qg}.

\section*{Acknowledgements}
Our work is supported by the TAMOP 4.2.1/B-09/1/KONV-2010-0007 and the
TAMOP 4.2.2/B-10/1-2010-0024 projects. The projects are implemented through the New
Hungary Development Plan co-financed by the European Social
Fund, and the European Regional Development Fund.

\end{document}